\documentclass[twoside]{article}
\usepackage{fleqn,espcrc2}

\title{Photoemission and the Origin of High Temperature Superconductivity}

\author{M. R. Norman\address{Materials Science
Division, Argonne National Laboratory, Argonne, IL  60439},
M. Randeria\address{Tata Institute of Fundamental Research, Mumbai
400005, India}, B. Jank\'o$^{\rm a}$,
and J. C. Campuzano$^{\rm a,}$\address{Department of Physics, University of
Illinois at Chicago, Chicago, IL  60607}}
       
\begin{document}

\begin{abstract}
The condensation energy can be shown to be a
moment of the change in the occupied part of the spectral
function when going from the normal to the superconducting state.  As a
consequence, there is a one to one correspondence between the energy gain
associated with forming the superconducting ground state,
and the dramatic changes seen in angle resolved photoemission spectra.
Some implications this observation has are offered.
\vspace{1pc}
\end{abstract}

\maketitle

In 1956, Chester published an interesting paper \cite{CHESTER} which dealt
with the the difference in energy between the normal and superconducting
states (the condensation energy).  For an isotope
coefficient of one-half, he demonstrated that the condensation energy could be
equated to the change in ion kinetic energy.
Historically, the paper did not play a major role,
since in the same year, the BCS theory of
superconductivity was being developed, solving the problem of classical
superconductors.

In this millenial year, though, we are faced with the unsolved problem of
high $T_c$ superconductivity.  Because of this, several of us
have invoked the name of Chester.  The hope is that by
directly focusing on the condensation energy, some light might be shed on
the solution to the high $T_c$ problem.  This is particularly relevant
if, as most of us suspect, the origin of high $T_c$ is associated directly
with electron-electron interactions.  In such a case, treating the pair glue
as an external object, as in the electron-phonon problem, could be 
misleading.  In essence, the electrons are gluing themselves together.  This
indicates that new ways of thinking may be important.

This has led a number of authors to concentrate on the change in various
response functions when going into the superconducting state.  This is
illustrated as follows:
\begin{eqnarray}
& & \Delta \rightarrow G,F \rightarrow R
\rightarrow F_N - F_S \rightarrow \Delta
\end{eqnarray}
The idea is to assume a non-zero superconducting order parameter, $\Delta$.
This leads
to changes in the normal Greens function, $G$, and to the creation of an
anomalous Greens function, the Gor'kov $F$
function.  This in turn causes changes in various two-particle response
functions, $R$: the dynamic spin susceptibility, the dielectric function,
the optical conductivity, etc.  This in turn leads to a change in the free
energy between the normal ($F_N$) and superconducting ($F_S$) states.
If the free energy is lowered, then a non-zero superconducting
order parameter is self-consistently stabilized.  Note that nowhere in this
argument does the question of the pair glue arise.  That is, superconductivity
is generated simply if the response function change is such as to lower the
free energy.

A number of theories with this philosophy have been advocated, each
focusing on a different response function.  One
due to Anderson and co-workers suggests that the
c-axis kinetic energy is lowered in the superconducting state \cite{ILT}.
This leads to a change in the c-axis optical
conductivity, which has received some experimental
support \cite{BASOV}.  A different suggestion
has been made in regards to the planar
conductivity \cite{HIRSCH}.

Turning to ``potential energy'' explanations, Leggett \cite{TONY} has
advocated that the energy savings comes from the density-density response
function.  Perhaps better known is the work of Scalapino and
White \cite{SW}, where a lowering of the exchange energy is suggested based
on a change in the spin-spin response function.  This idea was then connected
to the appearance below $T_c$ of the neutron resonance mode by Demler and
Zhang \cite{DZ}, which has also received experimental support \cite{DAI}.
In all of these cases, a particular part of the free energy is being singled
out, with the connection being made via a two particle
correlation function.

Here, a different approach is advocated \cite{COND}.  This is
illustrated as follows:
\begin{eqnarray}
& & \Delta \rightarrow G \rightarrow F_N - F_S \rightarrow \Delta
\end{eqnarray}
Note the simplified nature of this diagram relative to the first
one.  This argumentation is based on the following relation: \cite{MANY}
\begin{eqnarray}
\lefteqn{U_{N} - U_{S} =} \nonumber \\
\lefteqn{\sum_{\bf k} \int d\omega 
(\omega + \epsilon_k) f(\omega)
\left[A_{N}({\bf k},\omega) - A_{S}({\bf k},\omega)\right]}
\end{eqnarray}
where $U_N$ ($U_S$) is the internal energy of the normal (superconducting)
state, $A({\bf k},\omega)$ the single-particle spectral function, 
$f(\omega)$ the Fermi function, and $\epsilon_k$
the bare energy dispersion.
Eq.~3 is based on a reduced (single-band)
Hamiltonian with two particle interactions, and in principle can be generalized 
to the multi-band case by replacing the scalar quantities in this equation
by matrices in reciprocal lattice space.  It is easily demonstrated that
this equation using the BCS reduced Hamiltonian generates the BCS condensation
energy, $\frac{1}{2} N(0) \Delta^2$.

Note that the right hand side of Eq.~3 is a moment of the occupied part of the
single-particle spectral function $A$ ($A^-$).  There are strong
arguments that $A^-$ is being measured
by angle-resolved photoemission (ARPES) measurements in quasi-2D
systems \cite{MOHIT}.
As a consequence, we see that the high $T_c$ phenomenon is intimally
connected with the dramatic change in the photoemission lineshape when going
below $T_c$.

A useful decomposition, especially in regards to the various
theories mentioned above, is to break the right hand side of Eq.~3 up into
separate kinetic and potential energy pieces.  This is easily
implemented by rewriting $(\omega+\epsilon_k)$ as
$(2\epsilon_k) + (\omega-\epsilon_k)$, the first term being the kinetic
energy, the second the potential energy.  Therefore, changes in the kinetic
energy are associated with changes in the momentum distribution function
(which is related to the integrated ARPES spectral weight \cite{MOHIT}),
whereas on the Fermi surface, the potential energy contribution reduces to
the first moment of $A^-$.

This is easily illustrated for the case of BCS theory \cite{TINK}.
In this case, the potential energy
is lowered by $\Delta^2/V$, where $V$ is the pair potential, and the kinetic
energy increased by $\Delta^2/V - \frac{1}{2}N(0)\Delta^2$, the sum being
the BCS condensation energy.
The kinetic energy change is a consequence of the
broadening of the momentum distribution function by the BCS coherence
factors.  The potential energy change is
easily explained as well.  It is due to the difference in $\epsilon$
and $E=\sqrt{\epsilon^2+\Delta^2}$.  On the Fermi surface, this difference
is maximal, leading to a potential energy lowering of $\Delta/2$ (the factor
of $\frac{1}{2}$ coming from the coherence factors).  When integrated over
$\epsilon$, one then obtains $\Delta^2/V$.

There are two interesting points about the BCS example.  First, the transition
is potential energy driven (physically, this occurs because the ion terms
which actually drive the transition are absorbed into the effective
potential of the reduced Hamiltonian).  Second, the condensation energy
is confined to the vicinity of the Fermi surface by the coherence factors.
Note that although the ultraviolet cut-off (the Debye energy) enters the
individual kinetic and potential energy terms, it drops out of the net
term.

There are several reasons to believe that this BCS analogy may be misleading
in the high $T_c$ problem.  First, it assumes the existence of 
quasiparticles.  This can be contrasted with the high $T_c$ case, where 
although quasiparticle peaks exist below $T_c$, they do not exist
above \cite{KAMINSKI}.  That is, even though the superconducting state is
almost certainly a (superfluid) Fermi liquid, the normal state appears to be a
non Fermi liquid \cite{PWA}.  As a Fermi liquid does a better job of 
diagonalizing the kinetic energy than a non Fermi liquid, then one might 
conjecture that the kinetic energy is indeed lowered in the 
superconducting state despite the coherence factors.
Note that this does not violate the 
considerations of Chester \cite{COND}.  That is, although Chester \cite{CHESTER}
demonstrated that the potential energy of the electrons must be lowered and
the kinetic energy raised in the superconducting state, these refer to 
the potential and kinetic energy terms of the total Hamiltonian, not those of 
the reduced one.  Therefore, there is nothing that prevents the
kinetic energy of the reduced Hamiltonian from being lowered.

Second, the BCS condensation energy is confined to a narrow shell around the
Fermi surface.  For the
electron-electron case, though, we can anticipate that the whole Brillouin zone
could be affected.  This is corroborated by ARPES spectra, which do show changes
in the spectra even well away from the Fermi surface.  Moreover, ARPES data are
characterized by regions of the zone where the dispersion is weak.
These same regions of the zone are associated with the anomalous pseudogap seen
in underdoped cuprates.  Thinking about these regions of the
zone, even in the superconducting state, in terms of standard Fermi surface
based concepts may be misleading.

Eq.~3 suggests that the best way to get a handle on these issues is by
a detailed study of the change in the ARPES lineshape throughout the zone.
One objection is that the number being sought is small.  For
instance, Loram, based on specific heat data, estimates the condensation energy
to be only 3K per plane for optimal doped YBCO \cite{LORAM}.  On the other hand,
the relative contribution from a given ${\bf k}$ point is a different matter.
As noted above, in BCS theory, {\bf k} points on the Fermi surface yield a
contribution to the condensation energy of $\Delta/2$, which is a substantial
number.

A quick look at the data is sufficient to indicate potential
pitfalls, along with suggestions about what may be going on.
The $(\pi,0)$ point is singled out since it exhibits
the most dramatic lineshape change.  In the normal state, one has an
extremely
broad spectral peak, with a width of order the entire bandwidth.  In the
superconducting state, this gets dramatically rearranged into a sharp spectral
peak, followed at higher binding energy by a dip and hump.  By comparing these
two spectra, we get some idea about contributions of each spectral feature to
the condensation energy.

We begin by looking at the first moment ($\omega$) part of Eq.~3.
There will be a positive contribution
from the sharp spectral peak, which will be followed by a negative contribution
from the spectral dip.  Although the latter is smaller in weight than the
former, it is enhanced because of the $\omega$ weighting.  This is then
followed by the hump and subsequent tail region, and therein lies the rub.
Since this region is weighted by $\omega$, it is sensitive to how the
data are normalized.
Typically, ARPES data are normalized in such a way that the high energy tails
match, and thus to first approximation there is no tail contribution.  But,
if one assumes (on the Fermi surface) that the data are normalized by having
equal integrated weight, then typically the tails do not quite match.  The
resulting tail contribution to Eq.~3 can be quite large.
Similar considerations enter
for the kinetic ($\epsilon_k$) part of Eq.~3, where one deals with the change
in integrated area.
The message here is that since there are varying positive and negative
contributions to the condensation energy from any given spectrum, then the
normalization issue will have to be resolved before we can gain insight
from Eq.~3 based on ARPES data.

Still, a qualitative statement can be made
concerning the doping dependence of the condensation energy.  It is now well
known that the weight in the quasiparticle peak is dramatically suppressed
as the doping is reduced \cite{PRL99}.  In the context of Eq.~3, this implies
a decrease in the condensation energy with underdoping.  Moreover, since the
normal state in the underdoped case is gapped, one expects a further
reduction.  These two observations go a long way in explaining the
dramatic reduction in the condensation energy inferred from specific heat
data \cite{LORAM}.

To gain further insight,
we have studied \cite{COND} the so-called mode model \cite{PRL97},
developed to explain the lineshape change noted
above.  The idea is that the superconducting lineshape
is very similar to that expected for electrons interacting with a
collective mode.  Detailed analysis of the data vs. doping \cite{PRL99} has
verified
that the mode is almost certainly the resonance mode observed by
neutron scattering \cite{DAI}.  The model consists of a constant scattering
rate ($\Gamma$) in the normal state, which becomes gapped in the
superconducting state (this gap, which defines the spectral dip, is
equal to $\Delta$+$\Omega_0$ where $\Omega_0$ is the mode energy).
Of particular note, the quasiparticle peak is a consequence of a non-zero
$\Omega_0$.

The results for the condensation energy are surprising \cite{COND}.  Near the
Fermi surface, the kinetic
energy is lowered, and the potential energy raised.  The former occurs
because the formation of quasiparticle peaks has a larger effect on
sharpening the momentum distribution than the coherence factors have on
broadening it.  The potential energy lowering is
a more subtle matter.  The gap in Im$\Sigma$ causes Re$\Sigma$ by the
Kramers-Kronig relation to have a logarithmic behavior.  Because of this,
the normal and superconducting state spectra only asymptotically approach
one another.  This leads to a large negative tail contribution to the
first moment in Eq.~3, much larger than the positive contribution from the
quasiparticle peak.  As a result, the potential energy is raised.
The kinetic energy driven nature of the transition is surprising,
given the expectation that the neutron resonance mode should lead to a lowering
of the exchange energy \cite{DZ}.
This behavior, though, is sensitive to the size of $\Gamma$.  As
$\Gamma$ is reduced (the doping is increased), the normal state becomes more
Fermi-liquid like, and
one crosses over to the BCS limit, where the transition becomes potential
energy driven.

This behavior is reminiscent of a phase diagram for the cuprates recently
suggested by Phil Anderson \cite{PHIL}.  The potential energy is lowered
below a temperature $T^*$ due to the formation of the
pseudogap.
This line merges with $T_c$ on the overdoped side, and so the superconducting
transition is potential energy driven on this side.  On the underdoped side,
though, the transition is kinetic energy driven, since the potential energy
savings already occurs at $T^*$, and the additional kinetic energy savings is
driven by quasiparticle formation.  It is again important to note that despite
the presence of a large spectral gap in the underdoped case, quasiparticle
peak formation is still associated with $T_c$ \cite{PHENOM}.  Also, this
conjecture is consistent with Basov's results \cite{BASOV}, in that the sum
rule violation he reports (indicating a lowering of the kinetic energy) only
occurs on the underdoped side of the phase diagram.

In conclusion, we suggest that Eq.~3 will be very useful when thinking about
the origin of high $T_c$, and that ARPES data will play a major role in this
endeavor.

This work is supported by the U. S. Dept. of Energy, Basic Energy Sciences,
under Contract W-31-109-ENG-38, the National Science Foundation DMR 9624048,
and (M.~R.~) the Indian DST through a Swarnajayanti fellowship.


\begin{thebibliography}{9}

\bibitem{CHESTER}
G. V. Chester, Phys. Rev. {\bf 103}, 1693 (1956).

\bibitem{ILT}
S. Chakravarty, A. Sudbo, P. W. Anderson, and S. Strong, Science {\bf 261},
337 (1993).

\bibitem{BASOV}
D. N. Basov, {\it et al.}, Science {\bf 283}, 49 (1999).

\bibitem{HIRSCH}
J. E. Hirsch and F. Marsiglio, cond-mat/9908322.

\bibitem{TONY}
A. J. Leggett, Phys. Rev. Lett. {\bf 83}, 392 (1999).

\bibitem{SW}
D. J. Scalapino and S. R. White, Phys. Rev. B {\bf 58}, 8222 (1998).

\bibitem{DZ}
E. Demler and S.-C. Zhang, Nature {\bf 396}, 733 (1998).

\bibitem{DAI}
P. Dai, {\it et al.}, Science {\bf 284}, 1344 (1999).

\bibitem{COND} M. R. Norman, M. Randeria, B. Janko, and J. C. Campuzano,
cond-mat/9912043.

\bibitem{MANY}
L. P. Kadanoff and G. Baym, {\it Quantum Statistical Mechanics} (W. A. Benjamin,
New York, 1962), Ch. 2.

\bibitem{MOHIT}
M. Randeria, {\it et al.}, Phys. Rev. Lett. {\bf 74}, 4951 (1995).

\bibitem{TINK}
M. Tinkham, {\it Introduction to Superconductivity} (McGraw-Hill,
New York, 1975),p. 57.

\bibitem{KAMINSKI}
A. Kaminski, {\it et al.}, Phys. Rev. Lett. (Feb. 21, 2000).

\bibitem{PWA}
P. W. Anderson, {\it The Theory of Superconductivity in the High $T_c$
Cuprates} (Princeton Univ. Pr., Princeton, 1997).

\bibitem{LORAM}
J. W. Loram, K. A. Mirza, and J. R. Cooper, IRC Research Review (1998).

\bibitem{PRL99}
J. C. Campuzano, {\it et al.}, Phys. Rev. Lett. {\bf 83}. 3709 (1999).

\bibitem{PRL97}
M. R. Norman, {\it et al.}, Phys. Rev. Lett. {\bf 79}, 3506 (1997);
M. R. Norman and H. Ding, Phys. Rev. B {\bf 57}, R11089 (1998).

\bibitem{PHIL}
P. W. Anderson, private communication.

\bibitem{PHENOM}
M. R. Norman, M. Randeria, H. Ding, and J. C. Campuzano, Phys. Rev. B
{\bf 57}, R11093 (1998).

\end{thebibliography}
\end{document}